\documentclass{PoS}

\graphicspath{{./plots/}}
\usepackage{cancel}
\usepackage{amsmath}

\title{Inert Doublet Model signatures at future $e^+e^-$ colliders}

\ShortTitle{IDM signatures at future $e^+e^-$ colliders}

\author{\speaker{Dorota Soko{\l}owska}\\
       International Institute of Physics, Universidade Federal do Rio
  Grande do Norte,  Brasil\\
        E-mail: \email{dsokolowska@iip.ufrn.br}}

\author{Jan Kalinowski, Jan Klamka, Pawe{\l} Sopicki, Aleksander Filip \.Zarnecki\\
        Faculty of Physics, University of Warsaw\\      
        \email{jan.kalinowski@fuw.edu.pl},
        \email{j.klamka@student.uw.edu.pl},
        \email{pawel.sopicki@fuw.edu.pl},
        \email{filip.zarnecki@fuw.edu.pl}}

\author{Wojciech Kotlarski\\
        Institut f\"ur Kern- und Teilchenphysik, TU Dresden\\
        E-mail: \email{wojciech.kotlarski@tu-dresden.de}}

\author{Tania Robens\\
        Theoretical Physics Division, Rudjer Boskovic Institute, Zagreb\\
        E-mail: \email{trobens@irb.hr}}

\abstract{The Inert Doublet Model (IDM) is one of the simplest extensions of the Standard Model (SM), providing a dark matter candidate. It is a two Higgs doublet model with a discrete $Z_2$ symmetry, that prevents the scalars of the second doublet (inert scalars) from coupling to the SM fermions and makes the lightest of them stable. We study a large group of IDM scenarios, which are consistent with current constraints on direct detection, including the most recent bounds from the XENON1T experiment and relic density of dark matter, as well as with all collider and low-energy limits. We propose a set of benchmark points with different kinematic features, that promise detectable signals at future $e^+e^-$ colliders. Two inert scalar pair-production processes are considered, $e^+e^- \to H^+H^-$ and $e^+e^- \to AH$, followed by decays of $H^\pm$ and $A$ into final states which include the lightest and stable neutral scalar dark matter candidate $H$. Significance of the expected observations is studied for different benchmark models and different running scenarios, for centre-of-mass energies up to 3 TeV. Numerical results are presented for the signal signatures with two muons or an electron and a muon in the final state. For high mass scenarios, when the significance is too low for the leptonic signatures, the semi-leptonic signature can be used as the discovery channel.}

\FullConference{
European Physical Society Conference on High Energy Physics - EPS-HEP2019 -\\
			10-17 July, 2019\\
			Ghent, Belgium}

\begin{document}

\section{Inert Doublet Model}

The Inert Doublet Model (IDM) is one of the simplest extensions of the Standard Model (SM) which can
provide a dark matter (DM) candidate~\cite{Deshpande:1977rw,Cao:2007rm,Barbieri:2006dq}.
In addition to the SM doublet $\phi_S$, the scalar sector contains a second $SU(2)$ doublet, $\phi_D$. A discrete $Z_2$ symmetry is imposed on the model, with the following transformation rules for the fields: 
$\phi_S \rightarrow \phi_S, \; \; \phi_D \rightarrow - \phi_D, \;  \; \textrm{ SM} \rightarrow \textrm{ SM}$. Therefore, the additional doublet $\phi_D$, called dark or inert, is the only $Z_2$-odd field. The resulting $Z_2$-symmetric potential has a following form:

\vspace{-0.3cm}
\begin{equation*}\begin{array}{c}
V=-\frac{m_{11}^2}{2}(\phi_S^\dagger\phi_S)\!-\!\frac{m_{22}^2}{2}(\phi_D^\dagger\phi_D)+
\frac{\lambda_1}{2}(\phi_S^\dagger\phi_S)^2\! 
+\!\frac{\lambda_2}{2}(\phi_D^\dagger\phi_D)^2 \\
+\!\lambda_3(\phi_S^\dagger\phi_S)(\phi_D^\dagger\phi_D)\!
\!+\!\lambda_4(\phi_S^\dagger\phi_D)(\phi_D^\dagger\phi_S) +\frac{\lambda_5}{2}\left[(\phi_S^\dagger\phi_D)^2\!
+\!(\phi_D^\dagger\phi_S)^2\right].
\end{array}\label{pot}\end{equation*}

\vspace{-0.2cm}
Exact $Z_2$-symmetry implies that only  $\phi_S$ can acquire {a} non-zero vacuum expectation value ($v$). As a result the  scalar field{s}  {in $\phi_D$} do not mix with the SM-like fields {from} $\phi_S$, and {the} lightest particle of  {the} dark sector is stable. Therefore, there are five  physical scalars after electroweak symmetry breaking: the SM Higgs
boson $h$ and four dark scalars:  two neutral, $H$ and $A$, and two charged,  $H^\pm$. We here consider the case where the neutral particle $H$ is the lightest and therefore the DM candidate.

After electroweak symmetry breaking, the model contains seven free parameters. Agreement with the Higgs boson discovery and electroweak precision observables fixes the SM-like Higgs mass  {$m_h$} and $v$,  {and we are left with} five free parameters:
\vspace{-0.35cm}
$$m_H, \; m_A, \; m_{H^{\pm}}, \; \lambda_2, \; \lambda_{345}\,\equiv\,\lambda_3+\lambda_4+\lambda_5.$$

\begin{figure}[h]
\includegraphics[width=0.49\textwidth]{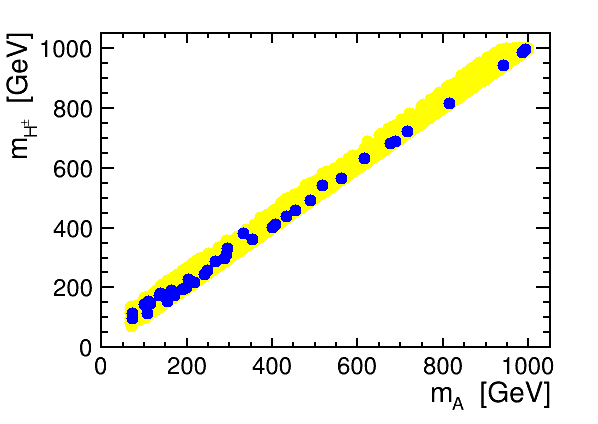}
\includegraphics[width=0.49\textwidth]{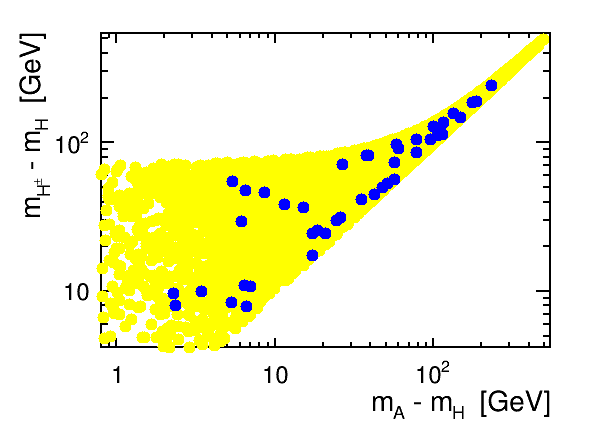}
\caption{Distribution of benchmark candidate points (yellow) and selected  benchmark points (blue) in the
  (m$_{A}$;m$_{H^\pm}$) plane (left) and in the
  (m$_{A} -\,$m$_{H}$;m$_{H^\pm} -\,$m$_{H}$) plane (right)~\cite{Kalinowski:2018ylg}.
}\label{fig:mass}
\end{figure}

\vspace{-0.35cm}
The model is a subject to various theoretical and experimental constraints, as described in detail in \cite{Kalinowski:2018ylg}. Among the theoretical constraints we include positivity of the potential, the condition to be in the inert vacuum and perturbative unitarity. Furthermore, we demand agreement with electroweak precision tests, electroweak gauge boson widths, no long-lived charged scalars, agreement with recasts of LEP and LHC searches, agreement with measurements of Higgs signal strengths and branching ratios, as well as null-results for additional scalar searches at the LHC. The DM candidate $H$ is constrained through results from direct and indirect detection experiments and measurements of DM relic density.
Based on these considerations, two sets of benchmark points (BPs) in agreement with all
theoretical and experimental constraints were proposed
in~\cite{Kalinowski:2018ylg}, covering different possible signatures
at $e^+e^-$ colliders, with masses of inert particles extending up to
1 TeV. Distributions of the scalar masses for these benchmark scenarios
are shown in Fig.~\ref{fig:mass}.

\section{Analysis strategy}

Prospects for the discovery of IDM scalars using leptonic channels at CLIC were described in
detail in~\cite{Kalinowski:2018kdn}. In this contribution we summarize
these results and extend them to the case of the ILC running at 250\,GeV and 500\,GeV centre-of-mass energy. We furthermore discuss the increased discovery reach we obtained from semi-leptonic channels at 3 TeV CLIC.

We consider the following tree-level production processes of inert scalars
at $e^+ e^-$ collisions: $ e^+e^- \to A~H$
and $e^+e^-\to H^+H^-$. The dark neutral scalar $A$
decays to a (real or virtual) $Z$ boson and the 
neutral scalar $H$, $A \rightarrow Z^{(\star)}H$, while the
charged boson $H^\pm$ decays predominantly to a (real or virtual) $W^\pm$ boson
and the neutral scalar $H$, $H^\pm \rightarrow {W^\pm}^{(\star)}H$. The lightest neutral particle $H$ is stable, and therefore escapes detection, contributing to the missing transverse energy ($\cancel{\it{E}}_{T}$).

\begin{figure}[htb]
\centerline{\includegraphics[width=0.75\textwidth]{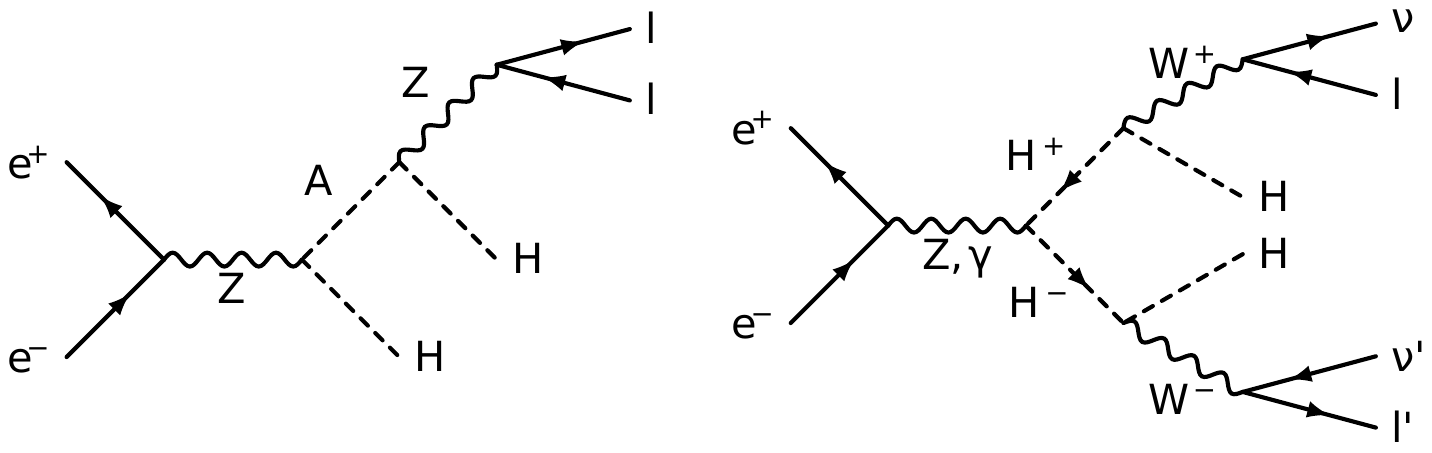}}
\caption{Signal Feynman diagrams for the considered production and
  decay process for:
{\sl (left)} neutral scalar production, $e^+e^- \to H A \to H H l l$,
and
{\sl (right)} charged scalar production, $e^+e^- \to H^+ H^- \to H H l l' \nu \nu'$.
}\label{fig:lepdiag}
\end{figure}

We first focus on leptonic decays of  $Z$ and $W^\pm$, as isolated leptons (electrons and muons) can be identified and
reconstructed with very high efficiency and purity. These processes lead to a signature of leptons and
missing transverse energy. The $\mu^+\mu^-$ final state predominantly arises from neutral scalar pair-production, while
the different flavour lepton pairs, $\mu^+ e^-$ and $e^+ \mu^-$ are considered a signature for the production of charged inert scalars, see
Fig.~\ref{fig:lepdiag}. However, during the simulation we do not constrain the intermediate states, but consider
all processes leading to $l l^{(')} + \cancel{\it{E}}_{T}$; both signatures include contributions from $AH$ and $H^+ H^-$ production. 
Especially processes with additional neutrinos, e.g. from $\tau$
(pair) production and their successive leptonic decays, have to be taken into account.
For the background, SM processes leading to charged lepton pair ($\mu^\pm e^\mp$ and $\mu^+ \mu^-$) and neutrinos (which constitute $\cancel{\it{E}}_{T}$) have been considered. As with the signal, topologies with intermediate $\tau$ leptons were also included.

\begin{figure}[htb]
\centerline{\includegraphics[width=0.4\textwidth]{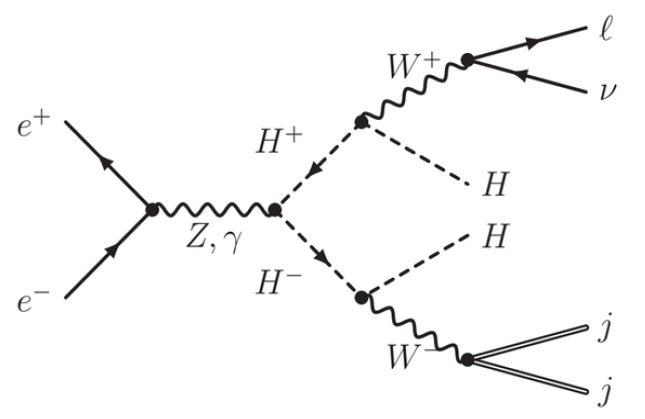}}
\caption{Signal Feynman diagram for the semi-leptonic decay process for charged scalar production:
$e^+e^- \to H^+ H^- \to H H l \nu jj$.
}\label{fig:diag}
\end{figure}

In the processes involving charged scalar pair production, one of the $W^\pm$ bosons can decay hadronically, leading to a semi-leptonic signature with $l \nu$ and two jets, as seen in Fig. \ref{fig:diag}. The hadronic branching ratio is significantly larger than the leptonic one, leading to a significant increase in the discovery reach. 

Details of the simulation, including cuts selection, are described in \cite{Kalinowski:2018kdn}. Signal and background samples were generated with WHizard
2.2.8~\cite{Kilian:2007gr}. Generator level cuts reflecting detector
acceptance for leptons and ISR photons were applied.
For the final selection of signal-like events, a multivariate analysis
is performed using a Boosted Decision Tree (BDT) classifier
\cite{Hocker:2007ht} with 8 input variables, both for the $HA$ and $H^+
H^-$ analysis. 
The BDT is trained using all accessible (at given energy) benchmark
points in a given category ($\mu^+\mu^-$ or $e^\pm \mu^\mp$ signature;
virtual or real $W$/$Z$). 

\section{Results for leptonic decay channels}

For both analysed signatures, the expected significance of the signal
is mainly related to the production cross-section for the considered channel.

\begin{figure}[htb]
\hspace{0.03\textwidth}
  \includegraphics[width=0.47\textwidth]{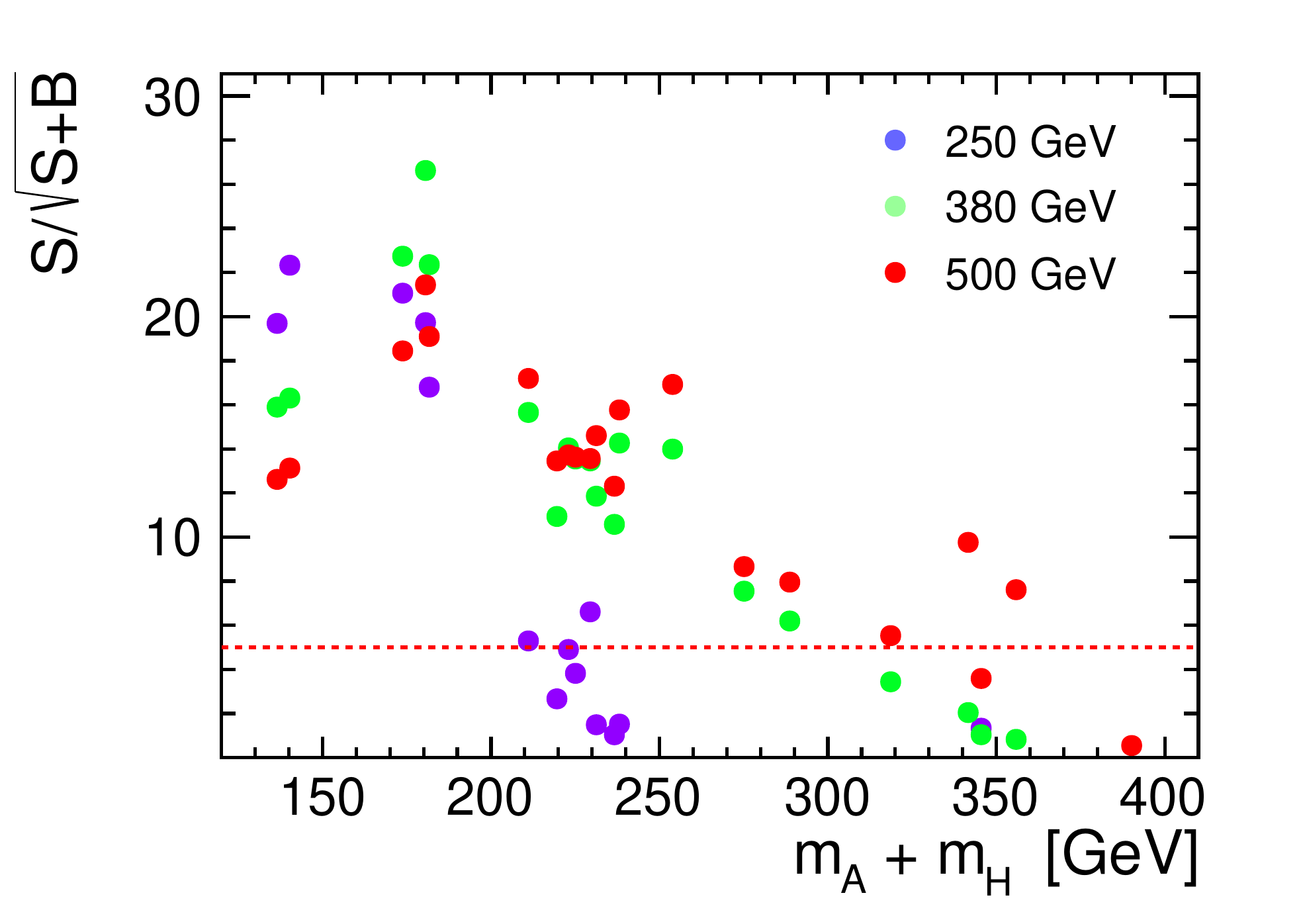}
  \includegraphics[width=0.47\textwidth]{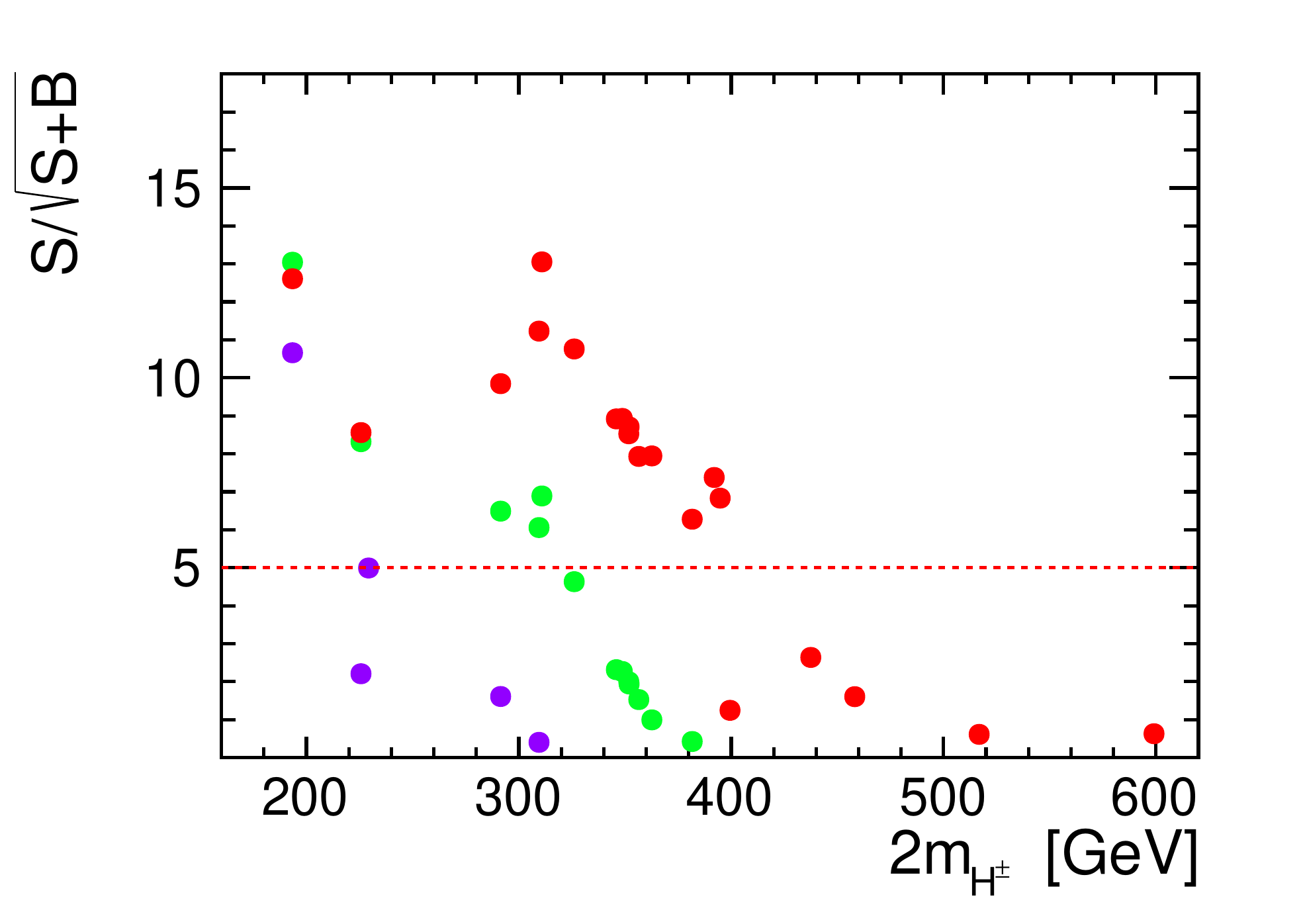}
\caption{Significance of the deviations from the Standard Model
  predictions, expected for 1\,ab$^{-1}$ of data collected at
  centre-of-mass energy of 250\,GeV, 380\,GeV and   500\,GeV, for:
  {\sl (left)} events with two muons in the final state ($\mu^+\mu^-$)
  as a function of the sum of neutral inert scalar masses and {\sl
    (right)} events with an electron and a muon in the final state
  ($e^+\mu^-$ or $e^-\mu^+$) as a function of twice the charged scalar
  mass.  
}\label{fig:lesig}
\end{figure}

\paragraph{Low $\sqrt{s}$ at ILC and CLIC} Fig. \ref{fig:lesig} shows the expected significance for low centre-of-mass energy stages of ILC and CLIC. In all cases, the integrated luminosity is set to 1\,ab$^{-1}$. For the centre-of-mass energies of 250\,GeV, 380\,GeV and 500\,GeV, the expected discovery reach of $e^+e^-$ colliders extends up to sum of neutral scalar masses of 220\,GeV, 300\,GeV
and  330\,GeV, respectively, and for charged scalar pair-production up
to charged scalar masses of 110\,GeV, 160\,GeV and  200\,GeV. Note, that as charged scalars are usually heavier than their neutral partners, fewer scenarios are accessible at a given energy in the electron-muon pair-production channel.

We usually do not observe strong dependence of the expected
significance on the mass splitting between inert scalars,
$m_{H^\pm} - m_H$ or $m_{A} - m_H$, within the considered range of
parameters. However, a possibility of significant contribution of cascade decays, $H^\pm \to W^\pm A \to W^\pm Z H$, which were not considered in the signal event selection, can reduce the signal, as is indeed the case with one of the benchmarks in the muon-electron channel (BP2 with $m_{H^\pm} = 112.8$ GeV).

\begin{figure}[htb]
 \hspace{0.03\textwidth}
  \includegraphics[width=0.47\textwidth]{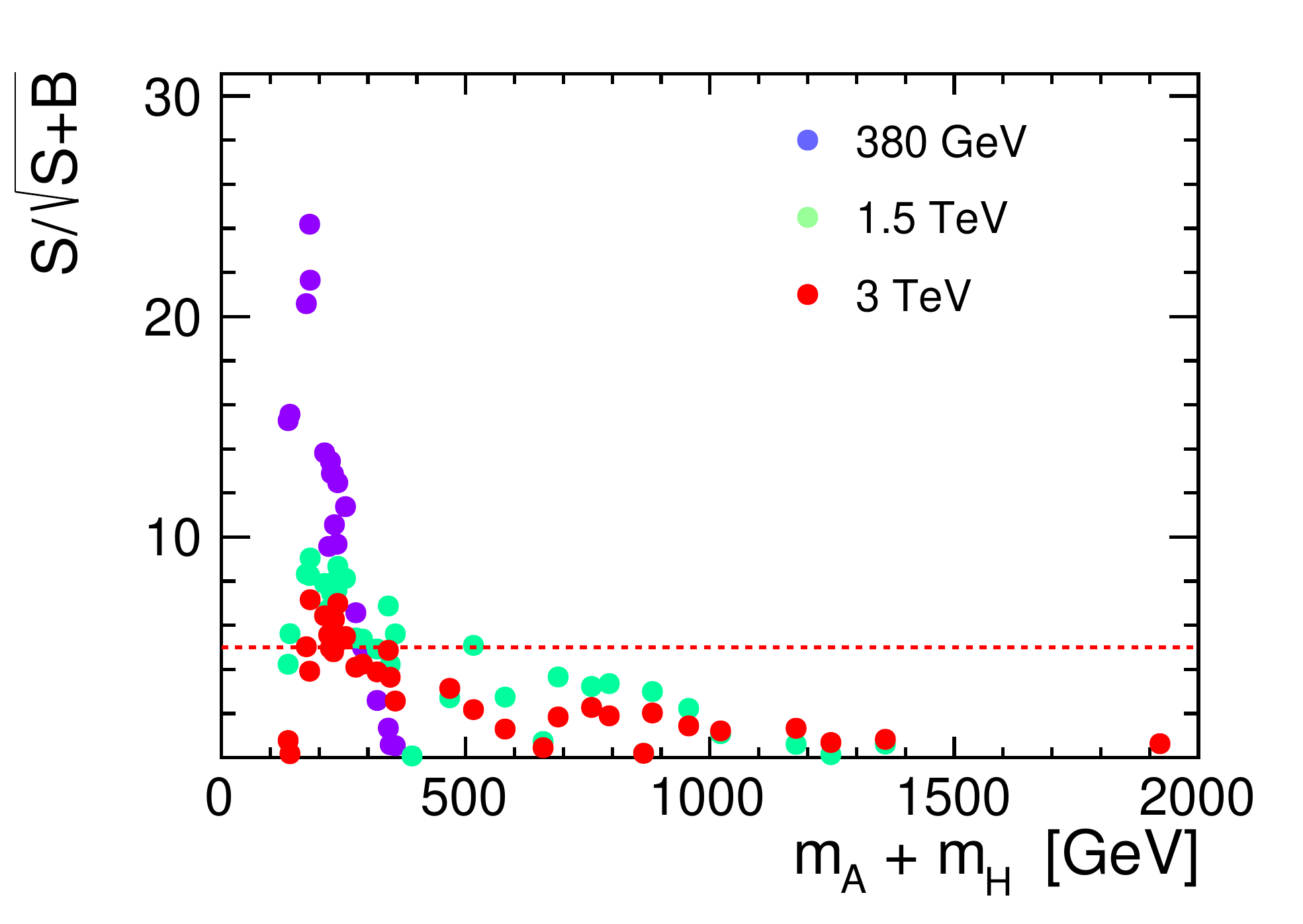}
  \includegraphics[width=0.47\textwidth]{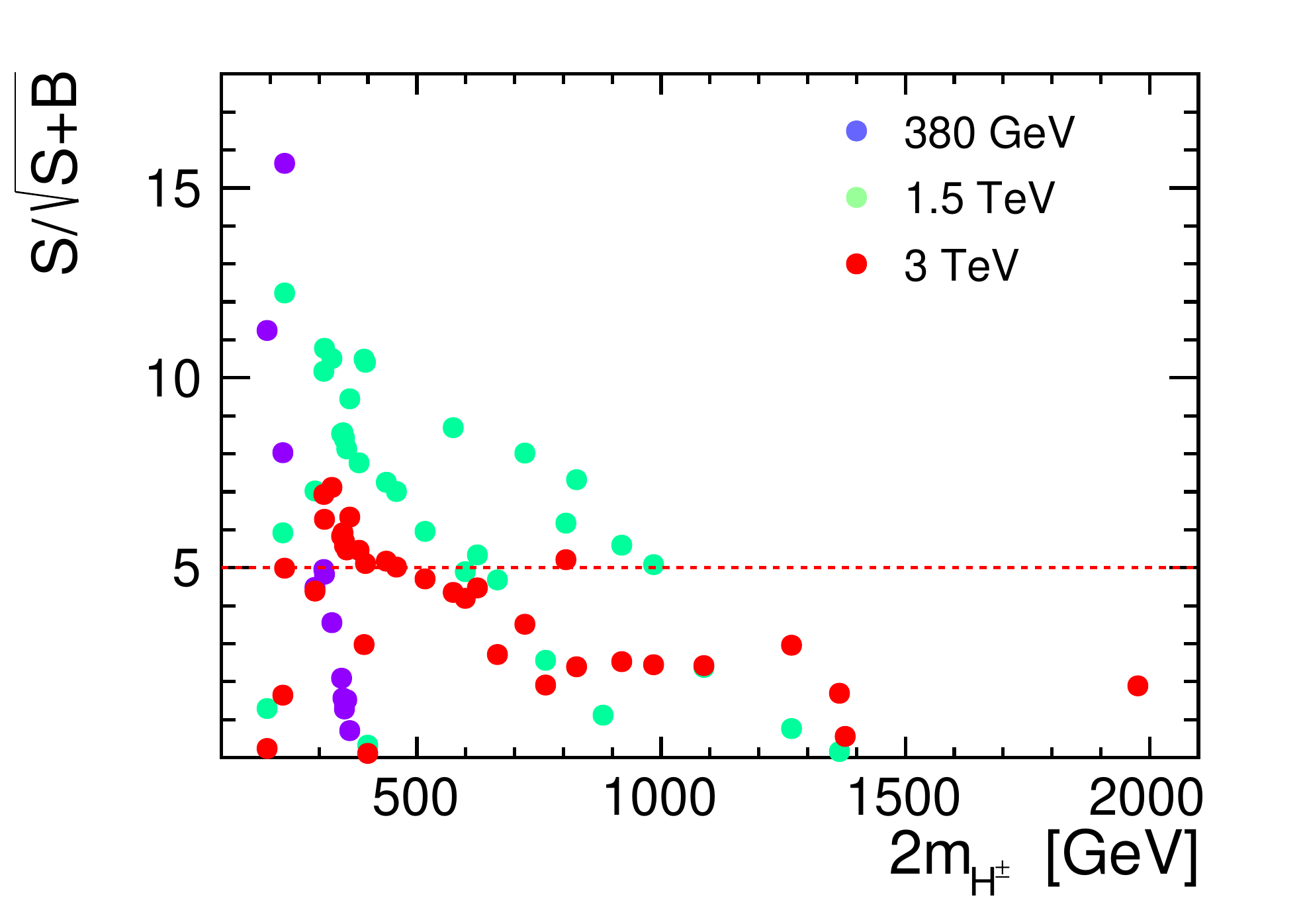}
  \caption{As in Fig.~\ref{fig:lesig} but for target integrated luminosity of CLIC: 1\,ab$^{-1}$ of data collected at 380\,GeV,
    2.5\,ab$^{-1}$ at 1.5\,TeV and 5\,ab$^{-1}$ at 3\,TeV.
  }\label{fig:hesig}
\end{figure}

\paragraph{High $\sqrt{s}$ at CLIC} If a scenario is not kinematically accessible at low energy stages of CLIC and ILC, it can be searched for at high-energy CLIC stages, with centre-of-mass energies 1.5 TeV and 3 TeV. After the kinematic threshold,
the signal production
cross-section for both considered channels decreases significantly with energy, much
faster than for the corresponding background. 
As shown in Fig. \ref{fig:hesig}, for 2.5\,ab$^{-1}$ of data and energy of 1.5\,TeV, only a moderate increase in discovery
reach is expected.
The neutral scalar pair-production in the $\mu^+ \mu^-$ channel can be discovered for $m_A + m_H < 450$\,GeV. A better reach was observed for charged scalar production in the $e^\pm \mu^\mp$
channels, leading to $m_{H^\pm} < 500$\,GeV.

For the scenarios considered
here, increasing the centre-of-mass energy to 3 TeV does not significantly improve the
sensitivity. Therefore, the observation of the inert scalar production in the leptonic
channels might be challenging at high-energy CLIC.

\section{Results for semi-leptonic decay channels}

\begin{figure}[htb]
\hspace{0.03\textwidth}
\centerline{  \includegraphics[width=0.47\textwidth]{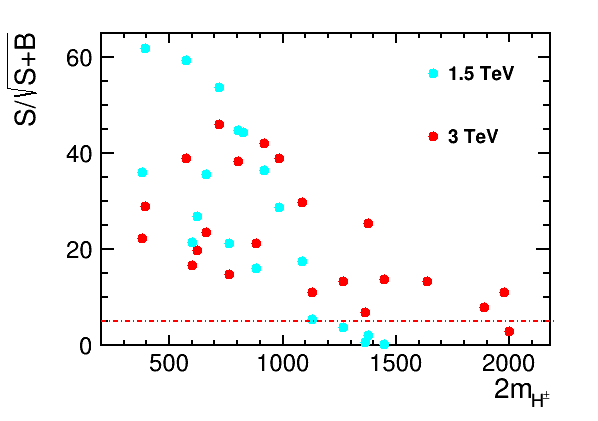}}
\caption{Significance of the deviations from the Standard Model
  predictions in the semi-leptonic channel as a function of twice the charged scalar
  mass expected for  2.5\,ab$^{-1}$ at 1.5\,TeV and 5\,ab$^{-1}$ at 3\,TeV.
  }\label{fig:slepsig}
\end{figure}

Due to the much larger branching ratios (28.6\% for $H^+H^- \to HH l^\pm \nu qq$, with $l= e, \mu$ compared to 2.3\% for $H^+H^- \to HH \mu^\pm e^\mp \nu \nu$) the expected number of $H^+H^-$  signal events in the semi-leptonic final state is over an order of magnitude larger than for the electron-muon signature. Considering a
similar scaling for the background processes (dominated by the $W^+W^-$
production), the expected significance of the observation in the semi-leptonic
channel  compared to purely leptonic one increases by a factor of 3-6, as presented in 
Fig. \ref{fig:slepsig}. For 2.5\,ab$^{-1}$ at 1.5\,TeV all proposed benchmarks with charged scalar masses up to 600 GeV can be discovered. Discovery reach is further expanded with increased energy and luminosity: for 5\,ab$^{-1}$ at 3\,TeV all benchmarks but one with charged scalar masses up to 1 TeV reach 5$\sigma$ significance.

\section{Conclusions}

A large part of the parameter space of the Inert Doublet Model can be tested at future $e^+ e^-$ colliders. Low mass scenarios can be observed with high significance in $\mu^+\mu^-$ or $e^\pm \mu^\mp$ channels already at the low energy stages of CLIC and ILC, up to $m_A + m_H \approx 330$~GeV and $m_{H^\pm} \approx 200$~GeV. The discovery reach is extended to $m_A + m_H \approx 500$~GeV and $m_{H^\pm} \approx 500$~GeV for $\sqrt{s} = 1.5$~TeV. There is no improvement in searches in leptonic channels with 3 TeV run at CLIC, however there is a significant increase in discovery reach while using semi-leptonic final states, providing 5$\sigma$ significance for inert scalar masses up to 1 TeV. 

\subsection*{Acknowledgements}

This contribution was supported by the National Science Centre, Poland, the
OPUS project under contract UMO-2017/25/B/ST2/00496 (2018-2021) and
 the HARMONIA project under contract UMO-2015/18/M/ST2/00518
 (2016-2019).

\end{document}